\documentclass[12pt]{article}
\usepackage[dvips]{graphics}
\usepackage{amstex}
\usepackage{amssymb}
\begin{document}

\begin{flushright}
INR-0951/97\\
August 1997
\end{flushright}

\vspace{0.5cm}

\begin{center}
{\large \bf Limits on leptonic photon interactions from SN1987a} 
\end{center}
\vspace{0.5cm}
\begin{center}
{\large  
S.N.Gninenko\footnote{E-mail address: Sergei.Gninenko\char 64 cern.ch}}
\end{center}

\begin{center}
{ Institute for Nuclear Research, Russian Academy of Sciences,\\ 
Moscow 117 312, Russia}  
\end{center}

\begin{abstract}
 If massless leptonic photons associated to 
electron, muon or tau leptonic number exist they would have been emitted from supernova 
1987a via the
annihilation process  $\nu\overline{\nu}\rightarrow\gamma_{l}\gamma_{l}$.\
By requiring that 
this process does not carry away most of the energy that can be radiated by the supernova 
 we obtain an upper limit on the leptonic photon coupling constants, 
$\alpha_{l} < 5.4 \times 10^{-11}$.\ Under certain assumptions about 
$\gamma_{l}$ - trapping conditions we find that the region of $\alpha_{l} = 10^{-8} - 10^{-7}$
might be of interest for an experimental search for leptonic photons.
\end{abstract}

\vspace{1.0cm}

Recently, a few papers devoted to leptonic photons, first discussed by Okun in 
ref. \cite{1}, have been published \cite {2,3,4,5}.
 
According to Okun \cite{1,2}, it is assumed that if leptonic quantum numbers  corresponding to the three lepton 
families ($\nu_{e},e$), ($\nu_{\mu},\mu$), ($\nu_{\tau},\tau$) are strictly
conserved these doublets could  carry leptonic charges, which are sources of electronic, muonic and tauonic massless photons, respectively.\ Okun considered the various consequences of this 
hypothesis and discussed possible experiments which might indicate the existence of 
these new vector particles.\ In particular,  he pointed out that the 
additional gauge symmetries connected with new leptonic photons would produce a problem of 
triangle anomalies, which need special attention.\      
To avoid this problem an anomaly-free scheme was suggested  in which a single leptonic 
photon $\gamma_{l}$ had the same coupling, but opposite sign  to the  ($\nu_{\mu},\mu$) and
($\nu_{\tau},\tau$)  doublets and was decoupled from ($\nu_{e},e$) \cite{2}.\ 
From the existing experimental data on muon magnetic moment upper limit $\alpha_{l}\le10^{-5}\times\alpha$ 
 could be obtained, where $\alpha$=1/137 is the fine structure constant
 ( hereafter we assume that $\alpha_{l}\equiv\alpha_{\mu}\equiv\alpha_{\tau}$).\\

The improved limit on $\alpha_{l}$ could be obtained from the analysis of high energy 
neutrino data.\ The idea is as follows \cite{2}.\ If leptonic photons exist 
they might result from the decay $\pi\rightarrow\mu\nu\gamma_{l}$ of charged pions copiously produced  
by high energy protons in the neutrino target.\ The $\gamma_{l}$'s 
 would penetrate the downstream shielding and would be observed in  neutrino
detectors via their conversion into $\mu^{+}\mu^{-}$ pairs in the Coulomb field of the 
target nuclei ( $\tau^{+}\tau^{-}$  pairs could also be produced, however the cross 
section for  their  production is 
suppressed by three orders of magnitude and in addition they are not easy identified).\  
The production rate of $\mu^{+}\mu^{-}$ pairs in the CERN SPS wide neutrino beam  was estimated in ref. \cite{6}.\ According to this paper the upper limit on $\alpha_{l}$ from the 
analysis of high energy neutrino data of the
experiments CHARM, NOMAD, or CHORUS is expected to be more than a factor of 10 better 
than the limit derived from the measurement of the muon (g$-$2).\\ 

   In a very recent publication Grifols and Masso pointed out that an upper limit of
1.7$\times 10^{-11}$ could be set on $\alpha_{l}$, based on the nucleosynthesis bound on the 
effective number of massless neutrino species, \cite{5}. However, in  ref. \cite{7}
it was argued that there is no cosmological bound on the leptonic fine structure 
constant $\alpha_{l}$ in the anomaly-free scheme described in the ref. \cite {2}.\\

 In this paper we show that experimental data from the supernova SN1987a
explosion in 1987 in the nearby Large Magellanic Cloud could be used to extract an 
upper limit on the coupling strength of leptonic photons, which is comparable with
the nucleosynthesis limit obtained in ref. \cite{5}.\\

The supernova SN1987a event has been extensively used to obtain limits on particle
 properties and interactions, see e.g. ref. \cite{8} and references therein.\ 
The basic and rather simple idea is as follows.\ According to supernova theories
most of the gravitational binding energy of the collapsed stellar core  is radiated in the form of neutrinos from its surface.\ The maximum possible gravitational energy release
is $E_{t}\approx4\times10^{53}erg$.\
From the observations of 
the neutrino pulse \cite{9} it is estimated that neutrino  carried out at least 
2$\times10^{53}erg$ of star energy.\ Taking into account that the neutrino diffusion time extends over 1 - 10 s leads to a limit for 
any additional energy drain $L$ from the supernova \cite{10}:

\begin{equation}
L \lesssim 2 \times 10^{53} erg/s
\end{equation}
     
Energy can be drained by any light particle produced in the SN1987a hot core
(where the temperature $T\sim60 MeV$) which might escape from the star without interactions
if its mean free path is longer than the core radius $R_{C}$ ( $R_{C}\sim 10^{6}cm$).
If a particle has a mean free path length smaller than $R_{C}$, it gets trapped and its 
subsequent thermalization can reduce the luminosity through diffusion out of the 
star core, as it happens for ordinary neutrinos.

Let us now consider the production of leptonic photons in the annihilation reaction 
$\nu\overline{\nu}\rightarrow\gamma_{l}\gamma_{l}$, where
 $\nu(\overline{\nu})$ denotes $\nu_{\mu}(\overline{\nu}_{\mu})$ or
$\nu_{\tau}(\overline{\nu}_{\tau})$ neutrinos. 
Since leptonic photons
are very weakly interacting with the core matter ( consisting mainly of electrons, neutrons or protons) we will assume that no trapping occures and they leave the star  without significant scattering.\ 

The energy emission rate ( energy per unit of volume and unit of time) for two interacting
distributions of neutrinos is given by the 
expression :
\begin{equation}
I(\nu\overline{\nu}\rightarrow\gamma_{l}\gamma_{l}) = 
\int \prod\limits_{i=1}^{2}\frac{d^{3}p_{i}}{(2\pi)^{3}2\epsilon_{i}}
\prod\limits_{j=1}^{2}\frac{d^{3}q_{j}}{(2\pi)^{3}2E_{j}}
\left|M\right|^{2}(2\pi)^4\delta^{4}(P_{i} - P_{j})(E_{1} + E_{2})f_{1}f_{2}
\end{equation}

where $\left|M\right|^{2}$ is the $\nu\overline{\nu}\rightarrow\gamma_{l}\gamma_{l}$ matrix element squared and summed over initial and final spins, $p_{i}(\epsilon_{i})$
is the neutrino momentum(energy), $q_{j}(E_{j})$
is the leptonic photon momentum(energy), $P_{i},P_{j}$ are the total initial and final 
momenta,  $f_{1},f_{2}$  are assumed to be the Fermi- Dirac statistics factors with zero 
chemical potential  ref. \cite{11}.\
For the initial neutrinos  $f_{1},f_{2}$ are given by

\begin{equation}
f_{i} = \frac{1}{e^{\epsilon_{i}/T} + 1}
\end{equation}

and  the neutrino number density $n_{\nu}$ (per $cm^{3}$) is given  by

\begin{equation}
n_{\nu} = \int\limits_{0}^{\infty} \frac{d^{3}p}{(2\pi)^{3}}f
\end{equation}

The SN1987a $\gamma_{l}$ - luminosity $L(\nu\overline{\nu}\rightarrow\gamma_{l}\gamma_{l})$ at fixed neutrino density can be written in the form  
  
\begin{equation}
\begin{split}
&L(\nu\overline{\nu}\rightarrow\gamma_{l}\gamma_{l}) = V \times 
I(\nu\overline{\nu}\rightarrow\gamma_{l}\gamma_{l})\\
&= V \times \int \frac{d^{3}p_{1}}{(2\pi)^{3}} f_{1}
\frac{d^{3}p_{2}}{(2\pi)^{3}}f_{2}
(\epsilon_{1} + \epsilon_{2}) \times 
\sigma(\nu\overline{\nu}\rightarrow\gamma_{l}\gamma_{l}) \times j
\end{split}
\end{equation}

where $V = 4/3 \pi R_{C}^{3}$, the
 factor $j = (p_{1}p_{2}/\epsilon_{1}
\epsilon_{2})v_{rel}$ is related to the flux
and the relative neutrino velocity  $v_{rel}$ =1 ( we assume that $m_{\nu} \ll T$). 
The cross section $\sigma(\nu\overline{\nu}\rightarrow\gamma_{l}\gamma_{l})$
can be written as 

\begin{equation}
\sigma(\nu\overline{\nu}\rightarrow\gamma_{l}\gamma_{l}) = 
\frac{4\pi \alpha_{l}^2}{s} (ln\frac{s}{k_{D}^{2}} - 1)
\end{equation}

 where $s$ is the square of the total energy of the two 
neutrinos in the centre of mass , $k_{D}$ is the Debye momentum defining
the range of the screened potential of leptonic charge in the neutrino -antineutrino plasma given by

\begin{equation}
k_{D}^{2} = \frac{\alpha_{l} n_{\nu}}{T}
\end{equation}

After numerical calculations we find that with good approximation equation (5) can be 
rewritten as  
\begin{equation}
L(\nu\overline{\nu}\rightarrow\gamma_{l}\gamma_{l}) = 
V \times \frac{2.8\alpha_{l}^{2}T^{5}}{\pi^{3}}
\times ln(\frac{10.3}{\alpha_{l}}) 
\end{equation}

Using $4/3 \pi R_{C}^{3}$ = $4\times10^{18}cm^{3}$, $T = 60 MeV$, $MeV^{5} = 3.2 \times 10^{47}
erg/cm^{3}/sec$
 one obtains:
\begin{equation}
L(\nu\overline{\nu}\rightarrow\gamma_{l}\gamma_{l}) \approx 2.4 \times 10^{75}\times \alpha^{2}_{l} erg/s  \lesssim 2 \times 10^{53} erg/s
\end{equation}
which would imply

\begin{equation}
\alpha_{l}\lesssim 9.1 \times 10^{-12}
\end{equation}

To estimate the uncertainty in the above limit we change the central temperature within 
the conventional range of $T \sim 30 MeV - 100 MeV$.  
This implies an uncertainty 
 factor of 6 in either direction in the limit in Eq. (10) and leads to the correct order of 
magnitude for the final limit on the coupling constant of leptonic photons 

\begin{equation}
\alpha_{l} < 5.4 \times 10^{-11}
\end{equation}

 Note that this limit is comparable with the nucleosynthesis limit obtained 
in  ref. \cite{5}. It is a few orders of magnitude  better than the limit 
obtained from the (g $-$ 2) experiment on muon 
magnetic moment or than the expected limit from the analysis of high energy neutrino data
 \cite{6}.

The result in Eq. (11) is obtained under the assumptions that the
leptonic photons are weakly interacting particles and that no trapping occurs.\
 To check that this is the case, we estimate the $\gamma_{l}$ mean free path 
$\lambda$ in the inner core as

\begin{equation}
\lambda= \frac{1}{n_{\nu} \sigma_{\nu} + n_{\gamma_{l}} \sigma_{\gamma_{l}}
+ n_{\mu} \sigma_{\mu}}
\end{equation}

where $n_{\nu}, n_{\gamma_{l}}$, and $n_{\mu}$ are  neutrino, $\gamma_{l}$ 
and muon densities, respectively
(it is assumed that $n_{\nu} \sim n_{\gamma_{l}} \sim n_{\mu}$),
 and $\sigma_{\nu}, \sigma_{\gamma_{l}}, \sigma_{\mu}$ are 
the cross sections for the processes 
$\gamma_{l}\nu(\overline{\nu})\rightarrow\gamma_{l}\nu(\overline{\nu})$, 
$\gamma_{l}\gamma_{l}\rightarrow\nu\overline{\nu}$, and
$\gamma_{l}\mu^{-}(\mu^{+})\rightarrow\gamma_{l}\mu^{-}(\mu^{+})$, 
respectively, which are given by 
formulae similar to Eq. (6).\ Here it is assumed that for $T \sim$ 60 MeV
the number density of muons in the core is the same order as muon neutrino concentration 
\cite{11}, while the $\tau$-lepton number density is negligible.\
 Since in the Okun model \cite{2} leptonic photons interact
 only with the second and third generations of fermions, other processes like 
$\gamma_{l}e\rightarrow \gamma_{l}e, \gamma_{l}p\rightarrow \gamma_{l}p, ...$
have much smaller cross sections than $\sigma_{\nu}, \sigma_{\gamma_{l}}$ 
and, hence, do not contribute to $\lambda$.    
 Using $\langle E_{\nu}\rangle \simeq \langle E_{\gamma_{l}}\rangle \simeq 3T$
 we have indeed $\lambda \gg R_{C}$ for the limit given in Eq. (11).
  
    If interactions of leptonic photons are strong enough, they will lead to
$\gamma_{l}$ trapping in the inner core and to $\gamma_{l}$ diffusion, rather 
than freely escaping.\ In this case, the relevant quantity is the diffusion 
time-scale associated with the mean free path $\lambda$ 

\begin{equation}
t_{diff} = \frac{3R_{C}^{2}}{\pi^{2} \lambda c}
\end{equation}

If the diffusion time $t_{diff}$ is of the order of 1 sec. or so, leptonic photons become 
 trapped and no longer  contribute effectively to the SN1987a luminosity.
After some simple calculations one obtains:

\begin{equation}
t_{diff} \approx 4.2\times 10^{12}\times \alpha_{l}^{2} \gtrsim 1
\end{equation}

which would naively imply 

\begin{equation}
\alpha_{l} \gtrsim 4.8\times 10^{-7}
\end{equation}

This limit is obtained at $T = 60 MeV$. Again, changing the temperature within
 $30 MeV - 100 MeV$  gives an additional 
uncertainty factor of 3 in either direction, since $n_{\nu} \sim T^{3}$ 
and diffusion time $t_{diff} \sim  n_{\nu} \cdot \alpha_{l}^{2}$.\ So, finaly

\begin{equation}
\alpha_{l} \gtrsim 1.6\times 10^{-7}
\end{equation}

Since this estimate  gives only a correct order of magnitude for the lower
limit on $\alpha_{l}$, we believe  
that region of $\alpha_{l} = 10^{-8} - 10^{-7}$ might  still be of interest 
for searching for leptonic photons in neutrino experiments.\ 
However,  more accurate calculations of 
$\gamma_{l}$-luminosity in the trapping conditions might provide a more 
stringent lower limit on $\alpha_{l}$, which would exclude any region of
$\alpha_{l}$ values between this lower limit and the  upper limit obtained from the of the muon (g$-$2) measurement.     
  
\vspace{0.5cm}

{\large \bf Acknowledgements}\\

The author is grateful to L.B.~Okun, N.V.~Krasnikov  and M.E.~Shaposhnikov  for useful discussions and 
remarks, to I.N.~Semeniouk for his help in numerical calculations, and  to
his colleagues from the NOMAD experiment at CERN for 
interesting  discussions and support.\ Special thanks to L.~DiLella and 
S.I.~Blinnikov for 
reading this manuscript and valuable comments.


\vspace{0.5cm}

\begin{thebibliography}{99}
\bibitem{1}
L.B.~Okun, Yad. Fiz. {\bf 10} (1969) 358 (in Russian); Sov. J. Nucl. Phys.~ 
{\bf 10} (1969) 206 (in English).
\bibitem{2}
L.~Okun, Phys. Lett. {\bf B382} (1996) 389. 
\bibitem{3}
S.~Blinnikov et al., Nucl. Phys. {\bf B458} (1996) 5264.
\bibitem{4}
A.~\c{C}ift\c{c}i et al., Phys. Lett. {\bf B355} (1995) 494.
\bibitem{5}
J.A.~Grifols and E.~Masso, Phys. Lett. {\bf B396} (1997) 201.
\bibitem{6}
V.A.~Ilin, L.B.~Okun and A.N.~Rosanov, preprint ITEP-ph-5-97, hep-ph/9707479.
\bibitem{7}
L.~Okun, Mod. Phys. Lett. {\bf A11} (1996) 3041.
\bibitem{8} 
A.~Dar, preprint hep-ph/9707501.
\bibitem{9}
K.~Hirata et al., Phys. Rev. Lett. {\bf 58} (1987) 1490;
R.~Bionta et al., Phys. Rev. Lett. {\bf 58} (1987) 1494;
E.N.~Alexeyev et al., Sov. JETP Lett. {\bf 45} (1988) 461.
\bibitem{10}
D.N.~Schramm, in: 1987 Inter.Symp. on ''Lepton and Photon Interactions at High Energies'',
eds. W.~Bartel and R.~Ruckl (Nopth-Holland, Amsterdam, 1988) p.471.
\bibitem{11}
I would like to thank S.~I.~Blinnikov for comments on this point.
\end{thebibliography}
\end{document}